\documentstyle[epsf,epsfig,wrapfig,here]{article}
\input psfig
\begin{document}           
\baselineskip=0.1666667in
\begin{quote} \raggedleft TAUP 2652-2000
\end{quote}
\vglue 0.5in
\begin{center}{\bf Remarks on Photon-Hadron Interactions}
\end{center} 
\begin{center}E. Comay 
\end{center}
 
\begin{center}
School of Physics and Astronomy \\
Raymond and Beverly Sackler Faculty \\
of Exact Sciences \\
Tel Aviv University \\
Tel Aviv 69978 \\
Israel
\end{center}
\vglue 0.5in
\vglue 0.5in
\noindent
\vglue 0.7in
\noindent
Abstract

   Theoretical aspects of VMD and related approaches to real
photon-hadron interaction are discussed. The work relies on
special relativity, properties of linearly polarized photons,
angular momentum conservation and relevant experiments. It is
explained why VMD and similar approaches should not be regarded
as part of a theory but, at most, as phenomenological models.
A further experiment pertaining to this issue is suggested.

\newpage
\noindent
{\bf 1. Introduction}
\vglue 0.33333in

   The discovery of the photon in the early years of the previous
century has identified it as a pure electromagnetic object. Many
years later it has been observed that hard photons interact with hadrons
in a manner which is akin to strong interactions and
looks independent of the electric charges of
the hadronic target. The following experimental results can be used as
an illustration of this
conclusion.

The cross section of hard photons scattered from a proton target is 
practically the same as that of a neutron one (see [1], pp. 292-293). 
Another kind of
data is the ratio between the numbers of
hadrons and leptons emitted
from a photon-proton interaction region. Here the number
of hadrons is greater by four orders of magnitude with respect
to the leptonic number. (This point is easily inferred from the
data discussed in [2], pp. 1567-1568 and from the table on
p. 323 of [1].) Thus, it is concluded that
"there is ample evidence which shows that the photon's hadronic 
structure plays a significant role in its interactions" (see the
abstract of [1]). 

   An approach attempting to explain experimental results of
photon-hadron interactions claims that a physical photon 
is composed of
a pure electromagnetic component {\em and} a hadronic one. 
According to this claim,
the wave function of a physical photon takes the form
\begin{equation}
\mid \gamma >\; = c_0\mid \gamma _0> + c_h \mid h>
\label{eq:GAMMA}
\end{equation}
where $\mid \gamma >$ denotes the wave function of a physical photon, 
$\mid \gamma _0>$
denotes the pure electromagnetic 
component of a physical photon and $\mid h>$ denotes 
its hadronic component. $c_0$ and $c_h$ are appropriate
numerical coefficients. Relation $(\!\!~\ref{eq:GAMMA})$ 
means that a real
photon fluctuates between a pure electromagnetic state and
a hadronic one. Moreover, this fluctuation is an {\em inherent 
property of the photon and is independent of 
its distance from the hadronic target}.
The relative time allotted to each state
is proportional to the absolute value of the square of
the corresponding coefficient of $(\!\!~\ref{eq:GAMMA})$. This approach
takes several ramifications, many of which are known as
Vector Meson Dominance (VMD) or Vector Dominance Models (VDM).
Here it is assumed that the hadronic part of 
$(\!\!~\ref{eq:GAMMA})$ is a neutral vector meson, which has the same
spin, parity and charge conjugation quantum numbers
as the photon (see eg. eq (2.1) on p. 271
of [1] and eq. (10.104) on p. 298 of [3]). A related claim states
that $\mid h>$, the hadronic part of $(\!\!~\ref{eq:GAMMA})$,
may belong to a larger set of hadronic states[4]. 
All these approaches are called below Photon's
Hadronic Structure Approaches (PHSA). The
present work examines critically the 
theoretical meaning of the common idea of PHSA,
which is manifested in $(\!\!~\ref{eq:GAMMA})$.

   A brief discussion of common properties and of differences
between the notions of a theory and a model is helpful for
a clarification of the main point of this work. The distinction
presented below between these notions 
should be regarded as a suggestion which is 
useful for the case discussed here. Obviously, other definitions
may be used, if they look helpful in other circumstances.

The following properties are common to a theory and to a model.
\begin{itemize}
\item[{A.}] Both provide a scheme leading to mathematical
formulas which describe experimental data. The scheme should
be mathematically selfconsistent.
\item[{B.}] Both are acceptable within an appropriate domain of
validity. (See [5] for a discussion of the notion of a validity 
domain of a theory.)
\item[{C.}] Both require a knowledge of certain constants which
are determined experimentally.
\end{itemize}

   On the other hand, the following properties distinguish
between a theory and a model.
\begin{itemize}
\item[{D.}] Within the corresponding validity domain, prediction
of a theory should be very precise whereas a model is acceptable
even if it yields just reasonably approximate predictions.
\item[{E.}] The physical constants used in a theory can be determined
by means of any set of
experiments, provided they are carried out within
the theory's 
validity domain. Another aspect of this point is that
in the case of a theory, after
fixing the required constants, one can apply {\em extrapolation},
into far regions, provided they are
included in the theory's validity domain. (Thus,
for example, after measuring the mass of a 
macroscopic body, one may use
Newtonian mechanics for all velocities which are much smaller
than the speed of light.) Contrary to this, a model is generally
useful only within a small domain where its constants have been
determined. In other words, a model is useful 
in cases where {\em interpolation} is applied
and deteriorates as it is extrapolated into far regions.
\item[{F.}] A model is tested by its practical benefit. 
If problems arise,
a model may be improved by an addition of certain corrections.
(Thus, for example, the nuclear liquid drop model is improved
by an addition of nuclear shell model terms, which account for nuclear
magic numbers.) By contrast, a theory 
is tested by its {\em correctness}.
In other words, a model is regarded as useful or not very useful
for certain applications whereas a physical theory can be 
{\em refuted} if it does not fit experimental data or well
established theories which have been confirmed by many experiments.
\end{itemize}

   It is explained in the rest of this paper why PHSA 
formulas belong to models
and do not constitute a part of a 
theory. This is probably the common belief of the
physical community as seen from the term VDM 
(Vector Dominance Models) and from its inclusion
in the phenomenological sections of PACS and of hep-ph@arXiv.org.

   The present work discusses only the theoretical side of PHSA.
On the other hand,
the problem of its usefulness as a model is beyond
the scope of the paper. In the second section it is shown that
PHSA is inconsistent with some 
well established theoretical results. Experiments
relevant to this matter are discussed in the third section.
Concluding remarks are the contents of the last section.
Expressions are written in units where $\hbar=c=1$. Energy-momentum
units are $MeV$ and $f^{-1}\simeq 197 MeV$. The cross section unit
is $mb = 0.1\,f^2$.

\newpage
\noindent
{\bf 2. Theoretical Problems of the Photon's Hadronic Structure
Approach}
\vglue 0.333333in

   Several theoretical difficulties of PHSA are pointed out here.

   Let us examine the implications of Lorentz transformations on the
coefficients $c_0$ and $c_h$ of $(\!\!~\ref{eq:GAMMA})$. For
this end, consider
Wigner's analysis of the Poincare group[6,7]. The analysis shows that
a massive particle can be regarded as an irreducible representation
of this group,
characterized by its self mass and spin. Massless particles, like the
photon, belong to a special case where spin is replaced
by helicity.

   This analysis is used in quantum field theory. It proves that
photons and hadrons are distinct objects. Since a quantum
mechanical state of a particle is characterized by the
eigenvalues of the self mass and the spin (or helicity),
one concludes that {\em every term} of its wave function should have
the same eigenvalues of these operators. It follows that 
$(\!\!~\ref{eq:GAMMA})$ {\em cannot} represent a quantum
mechanical state of a particle. This conclusion proves that PHSA is
inconsistent with relativistic quantum field theory.

   Another relativistic point is the behavior of $c_0$ and $c_h$
under Lorentz transformations.
It appears that in PHSA, it is assumed
that Lorentz transformations do alter these quantities, because
the hadronic part of soft photons is assumed to be negligible
(see [3], p. 298).
Thus, following this assumption, one does  not expect that
optical photons (or the black body radiation ones) interact strongly
with hadrons. The assumption that the 
relative size of the coefficients $c_0$ and $c_h$
of $(\!\!~\ref{eq:GAMMA})$ depend on the photon's energy is
denoted below as the energy dependence assumption.

   It is not clear how the energy dependence assumption is
embedded in a relativistic theory. Indeed, assume that one
measures energetic photons and finds that for $10\%$
of the time they interact like hadrons and for $90\%$ of the time
they interact like pure electromagnetic objects. Moreover,
as claimed by PHSA,
this ratio is an inherent property of the photon and is
independent of its proximity to an hadronic target. Hence,
relativity tells us that this ratio must be conserved for 
Lorentz transformations in general and for a Lorentz transformation
into a frame where the photon's energy is small,
in particular. This matter can be restated as follows. 
By their definitions, the
coefficients $c_0$ and $c_h$ of $(\!\!~\ref{eq:GAMMA})$
are the transition probabilities from a state of a physical
photon to that of a pure electromagnetic one and that of
a hadron, respectively. Now, "the transition probability
has an invariant physical sense" (see [6], top of p. 150). This
outcome is inconsistent with the energy dependence assumption.
Experimental aspects of this point are discussed in the next
section.

   Another issue is related to
transverse properties of photons. Thus, let us take
a linearly polarized photon moving parallel to the $z$-axis and
its electric field is parallel to the $x$-axis. Experiments 
measuring the interactions of such a photon with unpolarized target of 
protons are discussed below. Properties of linearly polarized  
photons clearly do not satisfy cylindrical symmetry around the
$z$-axis, because a rotation around this axis alters the direction
of its electric and magnetic fields. 
For photons of this kind, the vector potential
{\bf A} is parallel to the electric field. Hence,
since the interaction term of the electromagnetic Lagrangian
density is [8,9]
\begin{equation}
L_{int} = -j^\mu A_\mu,
\label{eq:JMUAMU}
\end{equation}
one finds that a linearly polarized photon
interacts with matter in a manner which breaks cylindrical symmetry
around the $z$-axis.

   Let us turn to the interaction of the assumed hadronic part
of this photon. Angular momentum conservation is utilized 
and calculations
carried out below show that, {\em under this restriction}, the 
assumed hadronic part of a photon interacts with an unpolarized
target in a manner which conserves cylindrical symmetry.
Special emphasis is put
on the $M$ values of the angular momenta, namely on their projection
on the $z$-axis.

   Due to angular momentum conservation, the angular momentum part
of the photon's hadronic state should be the same as that of the 
helicity of the (ordinary) electromagnetic state of the photon. 
Thus, since we have a linearly polarized photon, its
spin part has an equal amount of positive and
negative helicity (see [8], pp. 114-116; [9] pp. 273-275 
and [10]) and is written as a sum of two terms
\begin{equation}
\mid SM> = (\mid 11> + \mid 1-1>)/\sqrt 2.
\label{eq:SM}
\end{equation}
Here $S$ denotes spin and $M$ denotes its projection on the $z$-axis.
Due to angular momentum conservation, $(\!\!~\ref{eq:SM})$
describes also the spin state of the assumed hadronic part
of the photon.
Let us examine this state under a rotation by $\pi /2$ around the
$z$-axis. (This rotation exchanges the directions of the 
undulating electric and magnetic fields of the linearly polarized
photon.) Under this rotation, each term of
the wave function is multiplied
by $e^{-im\phi}$[11]. Thus, in the present case the corresponding
factor is $e^{\mp \pi /2} = \mp i$ and we have in the rotated frame
\begin{equation}
\mid SM>_{rot} = -i(\mid 11> - \mid 1-1>)/\sqrt 2.
\label{eq:SMR}
\end{equation}
Comparing $(\!\!~\ref{eq:SM})$ with $(\!\!~\ref{eq:SMR})$,
one realizes that, although each of the terms of
$(\!\!~\ref{eq:SM})$ varies only by a phase factor, {\em the relative
phase of the two terms changes sign}. This property means that 
cylindrical symmetry is broken if and only if
interference between the interactions of the two terms of
$(\!\!~\ref{eq:SM})$ does not vanish.
However, it is shown below that no such interference holds.
This result proves that the interaction of the hadronic
part of a photon with an unpolarized
target of protons is expected to conserve cylindrical symmetry.

   Three kinds of 
angular momenta are involved in the process: 
that of the assumed hadronic
part of the photon, $(\!\!~\ref{eq:SM})$, that of a proton at the
target (having $s=1/2$ and $m_s=\pm 1/2$)
and the spatial angular momentum between the incoming vector meson and
the proton participating in the interaction. Since the linear momentum
of the photon and of its assumed associated 
vector meson is parallel to the
$z$-axis, the projection of the spatial angular momentum on this
axis vanishes ({\bf r}$\times ${\bf p})$\cdot ${\bf p}=0.

   Let $M(1)$ and $M(-1)$ denote the overall $M$ value 
of the projectile-target system, pertaining to
the first and the second term on the right hand side of
$(\!\!~\ref{eq:SM})$, respectively. Following the discussion
carried out above, one sums the $M$ values 
of the three components and obtains
\begin{equation}
M(1) = 1 + (\pm 1/2),\;\;\;\;\;M(-1) = -1 + (\pm 1/2).
\label{eq:MM}
\end{equation}
Thus, since the Hamiltonian operator is a scalar in the
3-dimensional space,
interactions of the first term of $(\!\!~\ref{eq:SM})$ have
no common $M$ value with those of the second term. Therefore, no
interference between these interactions takes place and cylindrical
symmetry is expected to be conserved.

   This discussion shows that a pure electromagnetic linearly
polarized photon interacts with matter in a manner which breaks
cylindrical symmetry, as seen in $(\!\!~\ref{eq:JMUAMU})$.
On the other hand, the assumed hadronic part of such a photon 
conserves this symmetry. This result means that a transverse
information of the photon, namely - its linear polarization, 
{\em disappears} as the physical photon
fluctuates into a hadronic state. This property clearly reduces
the theoretical appeal of the VMD hypothesis. As
shown in the next section, it 
can also be used in an experimental test of its validity.

\vglue 0.666666in
\noindent
{\bf 3. Experimental Considerations}
\vglue 0.333333in

   Let us turn to experimental aspects of the topics discussed in
the previous section. First, the behavior of the coefficients
$c_0$ and $c_h$ of $(\!\!~\ref{eq:GAMMA})$ under Lorentz
transformations is examined. Two alternatives are discussed where
the energy dependence assumption holds or fails.

   Assume that the energy dependence assumption holds. This assumption
is probably made in order to settle problems of expected soft photon
interactions with hadrons, which otherwise emerge from the VMD
assumption. However, it leads to problems as one takes soft photons
and examines them in another inertial frame where these photons are
very energetic.

   Photon-photon interaction is probably
most suitable for this purpose, because,
unlike massive targets whose rest frame may look preferential, 
photons have no rest frame. Consider an
inertial frame $\Sigma$ and two sources of soft photon rays
(see fig. 1).
Here the photons interact electromagnetically and, as far as
the linearity of electrodynamics and Maxwell equations
hold, the photon-photon interaction
vanishes.

   Now, let us examine the process in another frame $\Sigma '$.
In $\Sigma $, $\Sigma '$ is seen moving parallel to the 
negative direction of the $y$-axis
and its velocity is not much smaller then the speed of light.
Hence, in $\Sigma '$, the photons emitted from $S_1$ and $S_2$
are very energetic. Now, if VMD
and $(\!\!~\ref{eq:GAMMA})$ hold then these photons should have
a hadronic part. Thus, in $\Sigma '$, the photon-photon interaction 
is expected to consist of two kinds of dynamical processes. The first one
is the pure electromagnetic process which is obtained from a Lorentz
transformation of what is found in $\Sigma$ and yields
a null quantity. The second process
is the hadron-hadron interaction which should take place under the
assumption examined here. This is a contradiction because the
percentage of events where photons 
interact and exchange energy-momentum should
be the same in all inertial frames.

   The second case is the ordinary quantum mechanical approach
where $c_0$ and $c_h$ of $(\!\!~\ref{eq:GAMMA})$ conserve
their absolute value under a Lorentz transformation (see [6],
top of p. 150). For examining this issue, let us take, for
example, the Compton scattering of 1 $MeV$ photon 
colliding with an electron of a hydrogen
atom. In this example,
calculations refer to the backwards direction $\theta = \pi$. The
Compton process is well known[12]. The angular dependence
and the energy of the emitted photon are 
obtained from the Compton relation
\begin{equation}
k_{out} = \frac {k_{in}}{1 + (2k_{in}/m)sin^2(\theta /2)}.
\label{eq:KOUT}
\end{equation}
Putting $k_{in}=1$ $MeV$, $m=0.511$ $MeV$ and $\theta = \pi$, 
one finds for the scattered photon
\begin{equation}
k_{out} \simeq 0.2 MeV.
\label{eq:KKOUT}
\end{equation}
The Compton unpolarized cross section is[12]
\begin{equation}
\frac {d\sigma }{d\Omega } = \frac {\alpha ^2}{2m^2}\left(\frac 
{k_{out}}{k_{in}}\right)^2 \left(\frac {k_{out}}{k_{in}} + 
\frac {k_{in}}{k_{out}} -
sin^2(\theta /2)\right),
\label{eq:DSIGMAC}
\end{equation}
where $\alpha \equiv e^2 \simeq 1/137$. In the present experiment, one
finds
\begin{equation}
\frac {d\sigma }{d\Omega }(\theta = \pi) \simeq 6.5 mb.
\label{eq:DDSIGMAC}
\end{equation}

   Let us turn to the photon-proton interaction. Here, the Compton
process $(\!\!~\ref{eq:DSIGMAC})$ can be ignored because the
proton/electron mass ratio is about 2000
and the cross-section is smaller by a factor of $1/4000000$. On the
other hand, if VMD-PHSA holds and
the photon has a hadronic component, then one expects
another process which is a meson-proton scattering. Since, in this
case, the proton's mass is 938 times that of the photon's
energy, one should have here a scattering process where the
photon's energy is (nearly) conserved.

   The effective radius of a meson-proton interaction region is
less than $\nolinebreak {10 f}$ 
and the photon's momentum is $1 MeV \simeq 1/197 f^{-1}$.
Hence, one finds that the spatial angular momentum practically
vanishes and, in a partial wave analysis, only the
S-wave contributes to the process.

   A crude estimate of the vector meson-proton cross section can be
obtained for the case discussed here from the data on $\pi$-proton
cross section[13]. Here one finds that in the low energy limit
\begin{equation}
\sigma \simeq 7 mb.
\label{eq:SIGMAPIP}
\end{equation}
Relying on a quark count, one concludes that a vector meson-proton
cross section is of the same order of magnitude as that of the
$\pi$-proton one $(\!\!~\ref{eq:SIGMAPIP})$. Hence, since we
have here an S-wave, the expected differential cross section is
obtained from a division of $(\!\!~\ref{eq:SIGMAPIP})$ by $4\pi $
\begin{equation}
\frac {d\sigma }{d\Omega } \simeq 0.6 mb.
\label{eq:DSIGMAPIP}
\end{equation}
Due to the assumption discussed here, where
$c_h$ of $(\!\!~\ref{eq:GAMMA})$ is not negligible, one 
compares $(\!\!~\ref{eq:DSIGMAPIP})$ with the backwards
Compton scattering differential cross section
$(\!\!~\ref{eq:DDSIGMAC})$. Thus, it is  found that if this version
of VMD-PHSA takes place, then a certain percentage
of the photons scattered backwards in an actual Compton
experiment should conserve the energy of the incoming photon
and violate the Compton relation $(\!\!~\ref{eq:KOUT})$ which
yields $(\!\!~\ref{eq:KKOUT})$. In other words, 
for $\theta = \pi $, the Compton
scattering yields outgoing photons whose energy is 0.2 $MeV$,
whereas the assumed
VMD-PHSA effect should yield 1 $MeV$ ones.

   To the best of the author's knowledge, this effect has never
been reported. Evidently, due to their energy difference,
a distinction between these kinds of
scattered photons can be easily made.
It is interesting to carry out such a test of
VMD-PHSA in an experiment dedicated to this problem.

   Another issue is the test of cylindrical symmetry in a
scattering process of linearly polarized photons on protons. As
shown in the previous section, electrodynamics breaks this
symmetry whereas the assumed vector meson is expected to 
interact with protons in a manner which conserves it. 
Related experiments have been carried out a long time
ago[14-16]. These experiments use linearly polarized photons 
and measure outgoing pions in
$\gamma p$ and $\gamma n$ collisions. The results
prove that cylindrical symmetry
is not conserved, contrary to what is expected from VMD.

\newpage
{\bf 4. Concluding Remarks}
\vglue 0.333333in

   This work examines theoretical aspects of VMD and related
approaches. It is shown above that VMD can be no more than
a phenomenological model. Evidently, 
if its merits are extended and it is
regarded as a part of a theory then it should stand
refutation tests. As a matter of fact, results of theoretical and
experimental tests show that VMD is inconsistent with some
well established theories and with experiments. A further
experiment dedicated to this issue can be carried out as
discussed in the third section.

   Another result of this work is that the hadronic features
of real photon-hadron interaction await theoretical interpretation.

\newpage
References:
\begin{itemize}
\item[{[1]}] T. H. Bauer, R. D. Spital, D. R. Yennie and F. M. Pipkin,
Rev. Mod. Phys. {\bf 50}, 261 (1978).
\item[{[2]}] H. Harari, Phys. Rev. {\bf 155}, 1565 (1967).
\item[{[3]}] H. Frauenfelder and E. M. Henley, {\em Subatomic
Physics}, (Prentice Hall, Englewood Cliffs, 1991). pp. 296-304.
\item[{[4]}] V. N. Gribov, Soviet Physics JETP, {\bf 30}, 709 (1970).
\item[{[5]}] F. Rohrlich, Classical Charged Particles,
(Addison-wesley, Reading mass, 1965). pp. 3-6.
\item[{[6]}] E. P. Wigner, Annals of Math., {\bf 40},
149 (1939). 
\item[{[7]}] S. S. Schweber, {\em An Introduction to Relativistic
Quantum Field Theory}, Harper \& Row, New York, 1964. pp. 44-53.
\item[{[8]}] L. D. Landau and E. M. Lifshitz, {\em The Classical
Theory of Fields} (Pergamon, Oxford, 1975). P. 71.
\item[{[9]}] J. D. Jackson, {\em Classical Electrodynamics} (John Wiley, 
New York,1975). p. 596. 
\item[{[10]}] S. Weinberg, {\em The Quantum Theory of Fields}
(Cambridge University Press, Cambridge, 1995). Vol 1, p. 74.
\item[{[11]}] A. de-Shalit and I. Talmi, {\em Nuclear Shell Theory}
(Academic Press, New York, 1963). P. 27.
\item[{[12]}] J. Bjorken and S. D. Drell, Relativistic
Quantum Mechanics (McGraw-Hill, New York, 1964) pp. 127-132.
\item[{[13]}] D. H. Perkins, {\em Introduction to High Energy Physics}
(Addison-Wesley, Menlo Park, CA, 1987). P. 113.
\item[{[14]}] D. Bellenger et al., Phys. Rev. Lett. {\bf 23}, 540
(1969).
\item[{[15]}] R. L. Anderson et al., Phys. Rev. {\bf D4}, 1937 (1971).
\item[{[16]}] D. J. Sherden et al., Phys. Rev. Lett. {\bf 30}, 1230
(1973).

\end{itemize}

\newpage
\noindent
Figure Captions

\noindent
Figure 1: 

   Two rays of light are emitted from sources $S_1$ and $S_2$
which are located at $x=\pm 1$, respectively. The
rays intersect at point $O$ which is embedded in the $(x,y)$ plane.

\newpage
\vglue 0.7in
\begin{figure}[htbp]
\centerline{\epsfxsize=10cm \epsfbox{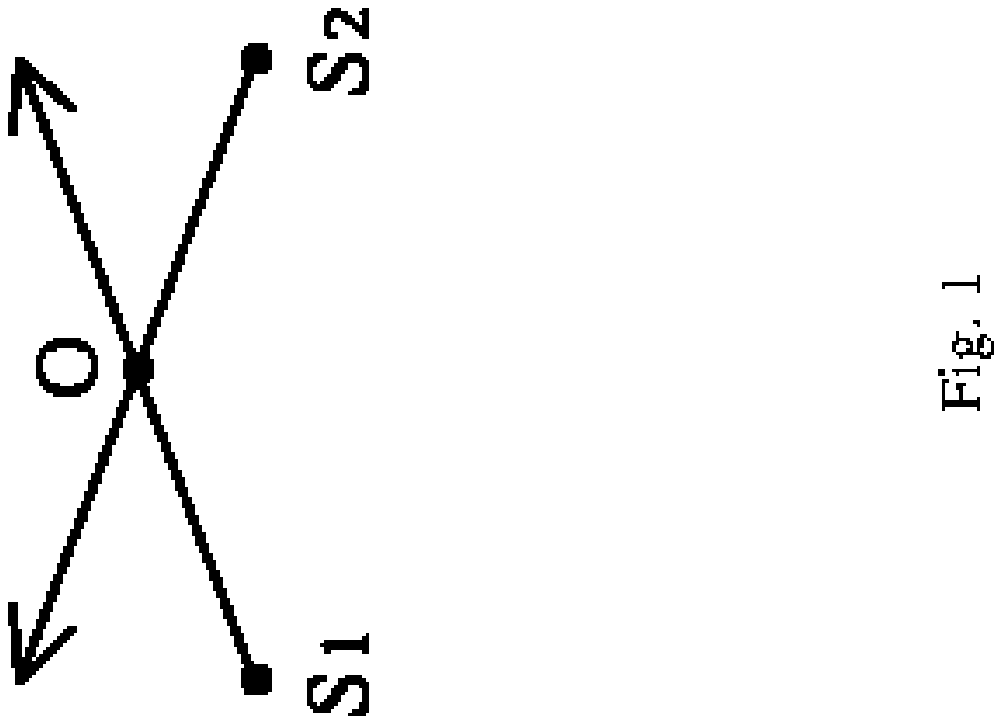}}

\end{figure}

\end{document}